\documentclass[aps,pra,twocolumn,showpacs,preprintnumbers,amsmath,amssymb]{revtex4-1}

 \usepackage{graphicx}

\begin{document}
\title{Thermalization of a strongly interacting closed spin system: From coherent many-body dynamics to a Fokker-Planck equation}
\author{C. Ates}
\author{J. P. Garrahan}
\author{I. Lesanovsky}
\affiliation{Midlands Ultracold Atom Research Centre (MUARC), School
of Physics and Astronomy, The University of Nottingham, Nottingham,
NG7 2RD, United Kingdom}


\begin{abstract}
Thermalization has been shown to occur in a number of closed quantum many-body systems, but the description of the actual thermalization dynamics is prohibitively complex. Here, we present a model - in one and two dimensions - for which we can analytically show that the evolution into thermal equilibrium is governed by a Fokker-Planck equation derived from the underlying quantum dynamics. Our approach does not rely on a formal distinction of weakly coupled bath and system degrees of freedom. The results show that transitions within narrow energy shells lead to a dynamics which is dominated by entropy and establishes detailed balance conditions that determine both the eventual equilibrium state and the non-equilibrium relaxation to it.
\end{abstract}
\maketitle
How a closed quantum system reaches a steady state in
which observables assume time-independent expectation values is both an intriguing and fundamental question in physics. When
such a steady state can be described by an equilibrium
thermodynamic ensemble we speak of ``thermalization''.
The study of closed quantum systems and their relaxation
behavior has recently received renewed interest \cite{Basko,Rigol,Eckstein,Popescu,olmu+:10,leol+:10,Cho,Srednicki2,genway,Huse,Biroli,Ponomarev,Cirac,Bunin,Yukalov} fuelled by ground breaking progress in preparing and manipulating ultracold atomic quantum gases \cite{Bloch}. Current experiments provide excellent thermal insulation from the environment and grant coherence times much longer than typical relaxation times. These approximately closed systems therefore constitute an ideal test bed for the investigation of relaxation phenomena and thermalization \cite{Greiner,Kinoshita,Hofferberth,Will,Trotzky}. In spite of this experimental progress it is fair to say that there is still a lack of understanding as to when and under what circumstances quantum systems prepared
far from thermal equilibrium relax to a steady state and when
the acquired equilibrium state is compatible with predictions
from statistical mechanics.

A currently very successful and insightful approach relates
thermalization to spectral properties of the many-body
Hamiltonian \cite{Deutsch,Srednicki,Goldstein,Reimann,Linden}. In thermalizing systems the expectation value of an observable $\hat{O}$ calculated in an eigenstate with energy $\epsilon$ coincides with the thermal average taken in the microcanonical ensemble with the same energy, i.e. $\langle\hat{O}\rangle_\mathrm{MC}(\epsilon)$. This eigenstate thermalization hypothesis (ETH) \cite{Deutsch,Srednicki} which
states that any individual many-body eigenstate contains a
thermal state is numerically verified for a number of systems
\cite{Rigol,Eckstein,Huse,Yukalov}. In practice any initial state $\left|\Psi\right>$ is a superposition of energy eigenstates $\left|\alpha\right>$. When these states are chosen to lie around a given energy $\epsilon$ shell relaxation to the microcanonical state will occur for long times as coherences between states of different energy wash out due to oscillating phase factors with incommensurate frequencies \cite{Rigol}. While the ETH is very useful to assess whether a system thermalises or not, the calculation of the actual thermalization dynamics requires the solution of the many-body Schr\"odinger equation. The complexity of this task grows exponentially with the number of degrees of freedom and even the most powerful numerical methods are restricted to small systems, low dimensions and/or short propagation times \cite{Trotzky}.

From classical statistical mechanics we know that relaxation is often described by an effective evolution equation, such as a Master or Fokker-Planck equation. It has the eventual equilibrium distribution as its stationary state due to the condition of detailed balance of its transition rates \cite{Gardiner}.  It is an intriguing question whether such Fokker-Planck equation can be actually also be found for a thermalising and closed many-body quantum system. If the answer is affirmative it will mean that expectation values of $\hat{O}$ can be calculated not only in the limit of infinite time, but also for intermediate times, and far from equilibrium, from a statistical ensemble.

\begin{figure*}
\centering
\includegraphics[width=1.8\columnwidth]{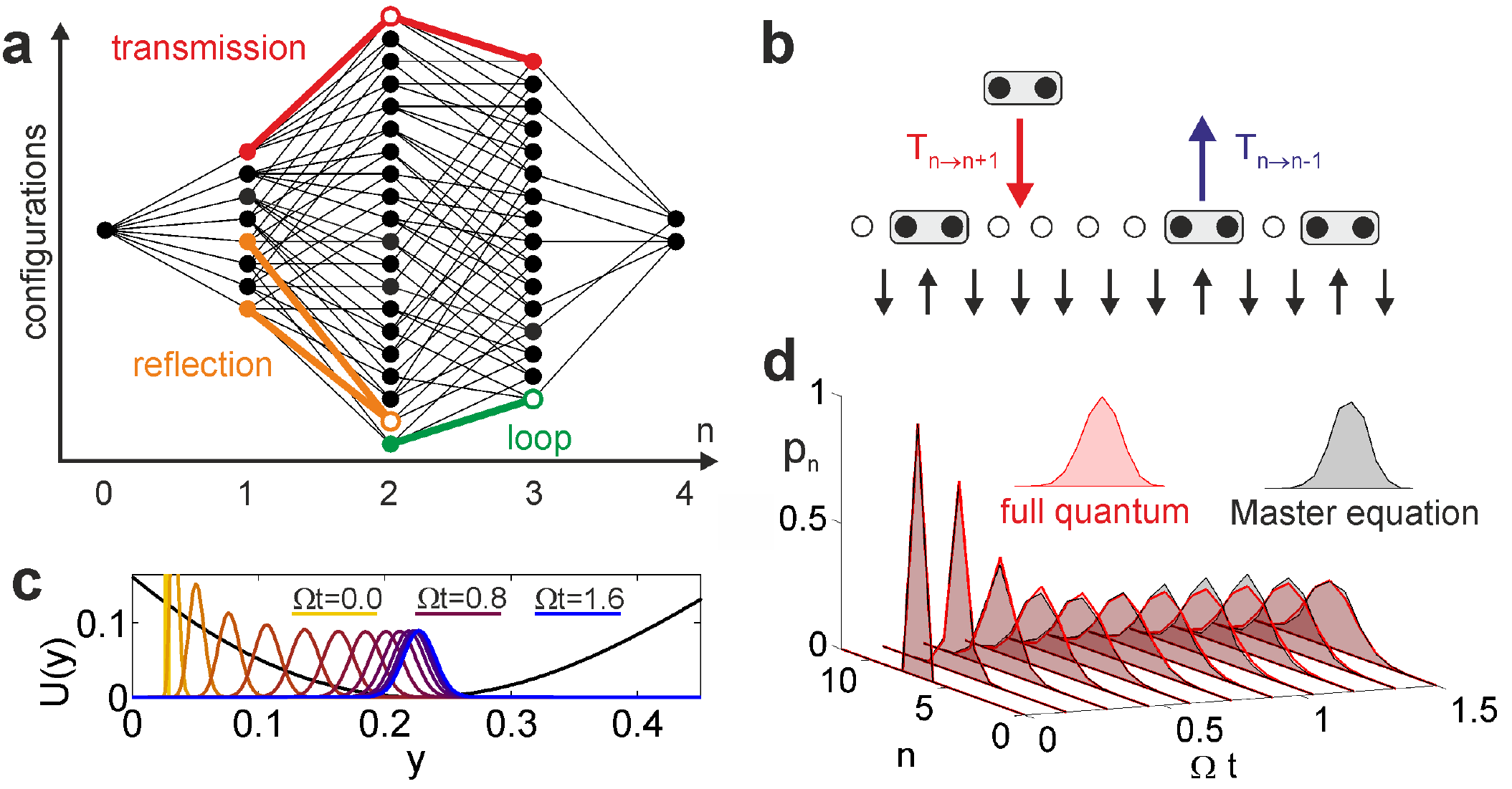}
\caption{{\bf a)} Graphical representation of the Hilbert space for a system with $L=8$ sites and $V\rightarrow\infty$. Each node represents a classical spin configuration with fixed excitation number $n$. The number of configurations with exciation number near $L/4$ grows exponentially in the system size. Links connecting the nodes represent transitions between spin configurations induced by the Hamiltonian. We obtain an effective evolution equation of the quantum state on the graph by a second order expansion of the von-Neuman-equation. Here only three fundamental moves can be identified: transmissions, reflections and loops. The solid colored circles shown represent beginning and end of such move while the hollow circles represent the intermediate state. In the thermodynamic limit loops constitute the dominant contribution. {\bf b)} Properties of the graph, such as transition coefficients between columns $T_{n\rightarrow n\pm1}$ can be analytically calculated by exploiting a connection to the partition function of a gas of hard core dimers in one dimension. An example for a spin configuration together with the corresponding dimer arrangement is shown. Deposition/removal of a dimer effectuates a move to the right/left on the graph. {\bf c)} The dynamics on the graph can be mapped to a one-dimensional Fokker-Planck equation describing diffusion of the probability distribution $\pi(y,t)$ in the potential $U(y)$. The variable $y$ is related to the excitation density $n/L$ by a transformation explained in the text. The shape of the potential is determined directly by the branching ratio of the graph a). An out-of-equilibrium distribution (yellow) relaxes into an equilibrium
state (blue) which is centered at the minimum of the potential. Different colors indicated different times in the interval
$0\leq\Omega t\leq 1.6$. {\bf d)} Temporal evolution of excitation number distribution
$p_n(t)$ for an initial state with $7$ excitations in
a system of $25$ sites. The plot shows data obtained from the
full quantum calculation (red) in comparison to the result
obtained from the Master Equation (\ref{master}). Both are in excellent agreement even for short times.
} \label{figure1}
\end{figure*}

Finding a general answer to this question is a formidable task. However, in order to show that this idea can work in principle we present a non-integrable spin model in which an effective Fokker-Planck equation can be analytically derived - a simple Ising-like spin-1/2 model on a lattice in a transverse magnetic field. Its Hamiltonian reads
\begin{equation}
\label{hamil}
H = H_{\Omega} + H_{V} =\Omega \sum_{k=1}^L \sigma_k^x + V \sum_{k=1}^L n_k n_{k+1},
\end{equation}
with the Pauli spin matrix $\sigma_k^x = \left( \left| \uparrow
\right> \left< \downarrow \right| + \left| \downarrow \right> \left<
\uparrow \right|  \right)_k$ and projector $n_k =\left( \left|
\uparrow \right> \left< \uparrow \right| \right)_k$ on the up-state on
lattice site $k$.  The coupling of the spins to the field is parameterized by $\Omega$. In contrast to the conventional Ising model the spin-spin interaction of strength $V$ is state-dependent, i.e.\ two neighboring spins only interact when both of them  are in the
up-state. We study this model because it exhibits generic features, such as non-integrability, even in one dimension, and local interactions leading to complex dynamics.  Furthermore, it is amenable to analytical treatment, and can be realised in experiment with ultracold lattice gases of laser-driven Rydberg atoms \cite{olmu+:10,leol+:10,welo+:08,le:11,Viteau}, polar molecules \cite{mibr+:06}, or with trapped ions \cite{poci:04,muli+:08}.

In the strongly interacting regime, $V \gg \Omega$ this system has been shown to thermalize \cite{olmu+:10,leol+:10}. Thermalization was demonstrated to become explicit in the distribution function $p_n(t)$ of the number of excitations (up-spins) which was shown to reach an equilibrium value for long times, i.e. $t\gg\Omega^{-1}$. $p_n(t)$ provides a meaningful measure of the degree of thermalization as it does not only capture local properties, e.g. the number of excitations
\begin{eqnarray}
  \langle N\rangle_t \equiv \sum^L_{k=1} \langle n_k\rangle_t=\sum_{n=0}^L n\, p_n(t),
\end{eqnarray}
but also encodes spatial correlations through
\begin{eqnarray}
  \langle N^2\rangle_t \equiv\sum^L_{k,m=1} \langle n_k n_m\rangle_t =\sum_{n=0}^L n^2\, p_n(t)
\end{eqnarray}
and higher moments.

To illuminate the thermalization process we consider the limit $V\rightarrow\infty$. Here the space of states with finite energy is spanned by all spin configurations in which neighboring excitations are absent. This space can be represented as the graph depicted in Fig. \ref{figure1}a (shown for $L=8$ for ease of visualization). The nodes of the graph correspond to classical spin configurations with fixed number, $n=0, \dots, L/2$, of (non-contiguous) excitations. Transitions between the states are driven by the Hamiltonian and shown as links connecting the nodes. For example the state $\left| \downarrow \uparrow \downarrow \uparrow \downarrow \downarrow \downarrow \downarrow \right>$ with $n=2$ excitations is directly coupled to the configuration $\left| \downarrow \uparrow \downarrow \uparrow \downarrow \downarrow \uparrow \downarrow \right>$ with $n=3$ up-spins but \emph{not} with $\left| \downarrow \uparrow \downarrow \downarrow \uparrow \downarrow \uparrow \downarrow \right>$ since both are not linked by just a single spin-flip operation. The quantum evolution of the system can be thought of as motion
within this graph, where the amplitudes of a quantum state
$\left|\Psi(t)\right>$ are related to the occupation of the
nodes.

The distribution function $p_n(t)$ is defined as
$p_n(t)=\mathrm{Tr}[P_n \rho(t)]$, where $P_n$ is a projection
operator onto the subspace spanned by all
states contained in the $n$-th column of the graph, see Fig.~\ref{figure1}a.  The density matrix for the whole closed system,
$\rho(t)=\left|\Psi(t)\right>\!\!\left<\Psi(t)\right|$, evolves according to the
von-Neumann equation ($\hbar = 1$)
\begin{eqnarray}
  \partial_t\rho(t)=-i\left[H,\rho(t)\right].
  \label{vonNeumann}
\end{eqnarray}
Starting from Eq.\ (\ref{vonNeumann}) it is possible to derive (see Supplementary material for details) for our spin problem a Master equation for the time evolution of the distribution function:
\begin{eqnarray}
\partial_t p_n(t) &=&
2\Omega^2\,t \;  \left[ T_{n+1 \to n} ~ p_{n+1}(t) + T_{n-1 \to n} ~ p_{n-1}(t)\right]
\nonumber \\
& &
-2\Omega^2\,t\; (T_{n \to n-1}+T_{n \to n+1}) ~ p_n(t)
\label{master}
\end{eqnarray}
The crucial insight that leads to this equation is that the complexity in the linkage of the graph, Fig.\ 1a, allows for a statistical treatment of the quantum dynamics. This becomes more accurate the larger the system size and for initial configurations that are located near the central region of the network, i.e., that are situated not too close to the left or right edge. Note that the latter condition is almost always satisfied for a randomly selected initial configuration.  Our approach bears similarities to the one presented in Ref. \cite{Cho}. The difference, however, is that we do not rely on a system-bath partitioning of the quantum system. The transition coefficients $T_{n \to n \pm 1}$ determine the probabilities for creating and annihilating excitations. They can be calculated analytically by exploiting the fact that the set of spin configurations that form the graph, Fig.\ref{figure1}a, are the same as those of hard-core dimers
on a one-dimensional lattice, Fig.\ref{figure1}b. The number of possibilities to remove/deposit one such dimer from a configuration with $n$ dimers is directly related to the typical number of links connecting a node in the $n$-th column of the graph.

Eq.\ (\ref{master}) is derived using a second order expansion where Hamiltonian Eq.\ (\ref{hamil}) induces only three kinds of transitions in the graph: {\em loop} transitions, {\em reflections}, and {\em transmissions}; see Fig.\ 1a.
In loop transitions the initial and final state is identical, while the intermediate state is a
configuration which differs by one excitation. Reflections are generalizations of loop transitions where the initial and final states have the same number $n$ of excitations but differ in their specific arrangement. For transmissions initial and final value of $n$ differ by two. Combinatoric arguments (see Supplementary material) imply that the contribution of loop transitions to the transition coefficients $T_{n \to n \pm 1}$ is by far the most dominant one. For instance, reflections between two \emph{randomly chosen} states are highly unlikely since due to the structure of the Hamiltonian both states must be identical except for the position of exactly one dimer. Explicitly, the transition coefficients $T_{n \to n \pm 1}$  are,
\begin{equation}
T_{n \to n-1} = n , \;\;\;
T_{n \to n+1} = \frac{(L - 2n -1)(L-2n)}{(L-n-1)}
\label{Ts}
\end{equation}
With these coefficients the Master equation (\ref{master}) obeys detailed balance, i.e., $p_n^{\rm{eq}} T_{n \to n+1} = p_{n+1}^{\rm{eq}} T_{n+1 \to n}$ for some equilibrium distribution $p_n^{\rm{eq}}$.  It can be easily verified that this distribution is given by \cite{footnote}
\begin{equation}
p_n^{\rm{eq}} = \left(\frac{2}{1 + \sqrt{5}} \right)^L \frac{L}{L-n}
{L-n \choose n} .
\end{equation}

In the continuum limit, $L,n \gg 1$ we can introduce the excitation density $x=n/L$ and  the Master Equation (\ref{master}) becomes a
Fokker-Planck equation (FPE) for the distribution function $p(x,t)$,
\begin{eqnarray}
\partial_t p(x,t)& =& -2\Omega^2 t  \left[ \partial_x F(x) - \frac{1}{2}
\partial_x^2 D(x) \right] p(x,t) \label{FP_equation} .
\end{eqnarray}
Notice that this FPE has both drift and diffusion coefficients, $F(x)=(1-5x+5x^2)/(1-x)$
and $D(x)=(1-3x+3x^2)/[(1-x)L]$, which are dependent on $x$.  Note also that while the FPE is local in time it depends explicitly on $t$ through the overall rate $2\Omega^2 t$ resulting in a typical time dependence $\sim\exp \left[- \lambda \Omega^2 t^2 \right]$ with $\lambda > 0$.  This is a direct consequence of the underlying coherent quantum evolution. The density dependent diffusion coefficient, $D(x)$, is a manifestation of the network structure of the state space in which the system moves according to the FPE.  Since $x$ is a continuous variable, $D(x)$ can be thought of as a metric.  All one-dimensional spaces are flat, so via a suitable coordinate transformation $y=y(x)$ we can bring the FPE to a form $\partial_t \pi(y,t) = - \Omega^2 t  [ \partial_y \tilde{F}(y)  - \tilde{D} \partial_y^2 ] \pi(y,t)$, where the new scaled diffusion coefficient is constant, $\tilde{D}=1$, and $\tilde{F}(y)=-\partial_y U(y)$ is a gradient force with the potential $U(y)$ (see Supplementary material).  In the $y$ representation relaxation corresponds to the approach to stationarity in the potential $U(y)$, see Fig.\ref{figure1}c.

\begin{figure}
\centering
\includegraphics[width=0.8\columnwidth]{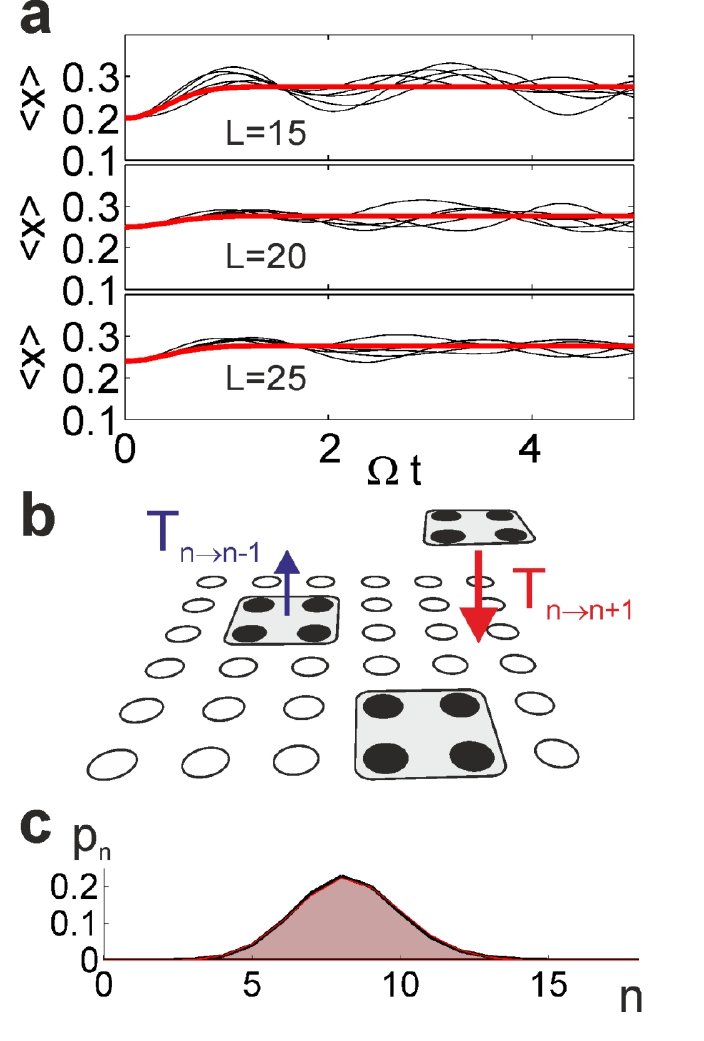}
\caption{{\bf a)} Finite size effects. Shown in black is the
evolution of the excitation fraction $\langle x\rangle=\langle
N\rangle/L$ for different system sizes, $L=15, 20, 25$. Each
panel shows data of the quantum calculation for $5$ initial states, randomly chosen from the manifold of $3$, $5$ and $7$ excitations. The red curve is the result obtained from the Fokker-Planck equation. The temporal fluctuations around the Fokker-Planck result decrease with increasing system size.
{\bf b)} The results also hold for higher dimensions, here two dimensions. The evolution can also here be mapped on a one-dimensional Fokker-Planck equation in excitation number space. The corresponding transition coefficients can be calculated from the partition function of a gas of hard squares. {\bf c)} Equilibrium distribution of the excitation number (black) for a two-dimensional $6\times 6$ lattice with periodic boundary conditions. The agreement with the numerically exact quantum result (red), here shown at $\Omega t=3.0$ for an initial state with $8$ excitations, is excellent. Note that the colormap is the same as in Fig. \ref{figure1}d.
} \label{figure2}
\end{figure}

A quantitative comparison between the diffusive dynamics predicted by the Master equation (\ref{master}) and the
exact quantum evolution is given in Fig.\ \ref{figure1}d.
Here we show the temporal evolution of the excitation number
distribution $p_n(t)$ for a system with
$L=25$ spins starting from a pure state with $7$
excitations. This state is
located in the central column of the configuration graph, cf.\ Fig.\ \ref{figure1}a,
where the assumptions that
entered in the derivation Eq.\ (\ref{master}) are best met.
That is, the initial conditions have to correspond to highly connected nodes of the graph, so that subsequent dynamics lead to a rapid spreading within the network of states.  If the initial state is near the edge of the graph, for example a state with no excitations, its low connectivity implies that the mixing is less rapid and coherent oscillations persist for longer times \cite{olmu+:10}.
Nevertheless, the vast majority of micro-states, and therefore most {\em randomly chosen} initial states, are located in the highly connected region of the graph and thus evolve according to the Master equation.
Indeed, Fig.\ \ref{figure1}d shows excellent agreement between the Master equation and the full quantum evolution for the relaxation of the distribution $p_n(t)$ from an initial state with fixed $n$ to the stationary state.  Both the thermal distribution
which is established for $\Omega t\geq 1.5$ and also the \emph{short time dynamics} is captured
very accurately by the Master equation (\ref{master}).
For small values of $L$ there are deviations from the thermal-like evolution.  These fluctuations are finite size effects which we expect to vanish for large enough systems: Fig.\ \ref{figure2}a shows that as anticipated temporal fluctuations become less pronounced with increasing $L$.

The obvious question is now how general these results are, for example, do they depend on dimensionality.  To address this we have also studied a two-dimensional version of the model (\ref{hamil}). Like in the
one-dimensional case the dynamics can be analysed with the help
of a graph. Configurations representing the nodes are given
by all possible arrangements of hard squares on a square lattice, see Fig.\ \ref{figure2}b. While the transition coefficients can no longer be calculated analytically, the numerically obtained Master equation for a $6 \times 6$ lattice with periodic boundary conditions shows a similar correspondence between the fully quantum evolution and the approximate Master equation as in the one-dimensional case, as illustrated in Fig.\ \ref{figure2}c.

In summary, we have shown for a spin model in one and two dimensions that the relaxation of a closed quantum many-body system to a thermal steady state is well approximated by a Master or Fokker-Planck equation. This is derived analytically from the full coherent dynamics, for the probability distribution of the thermalizing observables. The arguments that lead to such an evolution equation are the same that justify the eventual thermal steady state: the complex connectivity of the state space and the corresponding mixing effectuated by the Hamiltonian. This brings up the question of how the properties of the configuration network are related to the statistical complexity of the eigenstates that is central for the ETH. Clearly, both are connected as they stem from one and the same closed system Hamiltonian. Our model is physically relevant and yet generic - it is non-integrable, has local interactions and no disorder. Yet, an analytical analysis of the complex many-body dynamics is possible which is unusual for non-integrable systems. It represents a concrete example for how a thermalization dynamics can emerge from a purely quantum mechanical evolution without an explicit system-bath partitioning and/or postulated randomness.

We thank B. Olmos, L. Hackerm\"uller, P. Kr\"uger, M. M\"uller, S. Genway, and W. Li for comments on the paper.  This work was supported in part by EPSRC Grant No.\ EP/H024069/1
and Leverhulme Grant No.\ F/00114/BG.

%

\section{Supplementary Material}

For the interested reader we present here some technical details on the derivation of the Master equation as well as on the transformation of the Fokker-Planck equation to a form with constant diffusion term. 
\bigskip 

\noindent {\bf Basis states.}
In order to reflect the structure of the configuration network depicted in Fig. 1a in our calculations, we denote our micro-states (vertices of the network) in the from $ | n \, \mathcal{C}_n \rangle$.
Here $n$ denotes the number of up-spins and $\mathcal{C}_n$ their geometrical arrangement (configuration). The micro-states form a complete, orthonormal basis,
\begin{eqnarray}
\sum_{n \, \mathcal{C}_n} | n \, \mathcal{C}_n \rangle \langle n\, \mathcal{C}_n | &=& \mathbf{1} \nonumber \\
\langle n \, \mathcal{C}_n | m\, \mathcal{K}_m \rangle &=& \delta \left( n \, \mathcal{C}_n ,m \mathcal{K}_m \right).
\end{eqnarray}
Sometimes it will also be convenient to remind us about  the configuration energy $E = \langle n\, \mathcal{C}_n | H_{V} | n \, \mathcal{C}_n \rangle$ of a micro-state. In this case will denote it as $| n \, \mathcal{C}_n ; E \rangle$.


\bigskip

\noindent {\bf Sketch of the derivation of the classical Master equation.}
Using the integral form of the von-Neumann equation the time evolution of the expectation value $\langle \hat{O} \rangle_t$ of an observable is given by
\begin{equation}
\label{vneq}
\partial_t \langle \hat{O} \rangle_t = -\int_0^t \text{d}s \,
\text{Tr} \, \left\{  \hat{O} \left[ H_I(t), \left[ H_I(s), \rho(s) \right]
\right] \right\} ,
\end{equation}
where  $H_I(t) =
\mathbf{U}^{\dagger} H_{\Omega} \mathbf{U}$ is the interaction picture Hamiltonian obtained with the unitary transformation $\mathbf{U} = \exp [-it H_{V} ]$.
In writing eq.(\ref{vneq}) we have assumed that the initial state is such that the density matrix at $t=0$ commutes with $\hat{O}$. This is satisfied, if we assume that our initial state is a micro-state and our observable is the number of up-spins with fixed configuration energy $E$,
\begin{equation}
\hat{O} = \sum_{\mathcal{C}_n} {}^{\prime} | n\,
\mathcal{C}_n; E \rangle \langle n\, \mathcal{C}_n; E| .
\label{observable}
\end{equation}
The prime indicates that the summations has to be taken only over configurations with configuration energy $E$.
After inserting eq.\ (\ref{observable}) into eq.\ (\ref{vneq}), expanding the double commutator and using the cyclic property of the trace to interchange the density matrix to the rightmost side of each term, we perform three major steps to arrive at the final Master equation.

(i) Since our main interest is in the description of the time evolution of initial states that lie deep within our complex configuration network (i.e.\ that are connected to a huge number of other states), we assume that the system quickly loses its memory about its past. Our first approximation is, therefore, to replace $\rho(s) \to \rho(t)$ in eq.\ (\ref{vneq}), which is effectively a second order expansion of the von-Neumann equation reminiscent of the Redfied equation \cite{brpe:02}.

(ii) We only keep the statistically dominant loop transitions in our description (see next section for the justification of this step and ref.\ \cite{olmu+:10} for a numerical example). This amounts to only keeping the diagonal elements of the squared Hamiltonian $H_I$.  After these replacements we perform the time integrals that yield functions of the form $r(t) = \sin \left[\left( E^{\prime} - E  \right) t \right] / \left( E^{\prime} - E \right)$. Since we restrict ourselves to the dynamics within one energy shell ($E^{\prime} = E$) this function simplifies to $r(t) \to t$.

(iii) Using the fact that the Hamiltonian only couples states that differ in excitation number by one, we can perform the summation over the excitation number index. In our final approximation we replace the remaining matrix elements of $H_I^2$  by $\Omega^2$ times the mean connectivity of the $n$-manifolds (see below for details) and arrive at the Master equation presented in our main manuscript.


\bigskip

\noindent {\bf Relative importance of the three transition types.}
In the derivation of the Master equation we only take loop-transitions into account (c.f. Fig. 1a). The neglect of the other transition types for $n \gg 1$ is justified as follows: If we randomly pick a state with $n$ excitations the number of \emph{loop transitions} is simply given by the number of direct connections of a chosen
mirco-state to states with one more or less up-spins. 
This number is on the order of magnitude of the system size $L$. This has to be contrasted with the situation faced when regarding \emph{reflections}. If one randomly picks an initial and a distinct
final micro-state within a manifold having $1 < n < L/2$ excitations, these states will most probably not be connected at all by a reflective transition, since the spin-flip term of the Hamiltonian can only couple two
states that differ by the position of one excitation. The probability that two states are linked by a reflection can be assessed  by dividing the number of configuration pairs that exactly differ by the position of one up-spin
(which is on the order of $L$) by the total number of configurations $\nu_n$ with $n$-excitations (that for $n \gg 1$ grows exponentially with system size). It is vanishingly small for  $1 < n < L/2$. By a similar argument one can also see that \emph{transmissions} are statistically insignificant as compared to loop transitions.


\bigskip

\noindent {\bf Calculation of graph connectivity.}
In order to obtain the mean connectivity between manifolds with $n-1$ and $n$ up-spins we start by determining the \emph{total} number of allowed transitions from states with $n$ to states with $n -1$ excitations,
\begin{equation}
c_{n \to n-1} = \langle n -1  | \sum_k \sigma_k^- | n \rangle ,
\end{equation}
where $\sigma_k^-$ is the spin lowering operator at site $k$ and $|n\rangle = \sum_{\mathcal{C}_n} | n \, \mathcal{C}_n \rangle$. The calculation is most easily done by using a spin-coherent state $| \xi \rangle$ introduced in ref.\ \cite{le:11},
\begin{eqnarray}
| \xi \rangle &=& \prod_k^L (1 - \xi m_{k-1} \sigma_k^+ m_{k+1}) | 0 \rangle \nonumber \\
 &=& | 0 \rangle - \xi | 1 \rangle + \xi^2 | 2 \rangle - \dots
\label{rk_wavefunc}
\end{eqnarray}
with $m_k = \mathbf{1} - n_k$. This state is a weighted sum over all allowed micro-states in the $E=0$ shell. It is normalized to the partition function $\Xi (z) $ of a hard-dimer gas with fugacity $z = \xi^2$.

Interestingly, expectation values of off-diagonal operators taken with the state $| \xi \rangle$ can be expressed by expectation values of diagonal operators. In particular we have $\langle \xi | \sum_k \sigma_k^- | \xi \rangle = -\xi^{-1} \langle \xi | \sum_k n_k | \xi \rangle$ with
\begin{equation}
\langle \xi | n_k | \xi \rangle = \frac{1}{2} \left( 1 - \frac{1}{\sqrt{1+4z}} \right) \Xi (z)
\label{density}
\end{equation}
Using these relations we can define a generating function $\Lambda (z)$ for the coefficients $c_{n \to n-1}$
\begin{equation}
\Lambda (z) = \sum_k \langle \xi | n_k | \xi \rangle = \sum_n z^n c_{n \to n-1}
\end{equation}
so that
\begin{equation}
c_{n \to n-1} = \left. \frac{1}{n!} \frac{\partial^n}{\partial z^n} \Lambda (z) \right|_{z=0} =\frac{L}{(n-1)!} \prod_{j=n+1}^{2n-1} \, ( L - j) .
\label{cproduct}
\end{equation}
The typical number of connections (i.e. possible transitions) $T_{n \to n-1} $ that \emph{a randomly chosen state} from the manifold with $n$ excitations has to the $n-1$ -manifold can then be obtained by dividing (\ref{cproduct}) by the total number $\nu_n$ of states with $n$ excitations generated by $\Xi (z)$,
\begin{equation}
\nu_n = \left. \frac{1}{n!} \frac{\partial^n}{\partial z^n} \Xi (z) \right|_{z=0} = \frac{L}{n!} \prod_{j=n+1}^{2n-1} \, (L-j) .
\label{numconf}
\end{equation}
Similarly all other coefficients of the Master equation can be obtained.


\bigskip

\noindent {\bf Fokker-Planck equation and transformation to constant diffusion term.}
Our aim is to transform the Fokker Planck equation such, that the diffusive term becomes constant. In general such a transformation reads
\begin{equation}
p(x) \text{d} x = \pi(y) \text{d}y =  \pi(y) \frac{\text{d} y}{\text{d} x} \text{d} x \equiv  J \, \pi(y) \text{d} x
\label{trafo_general}
\end{equation}
with the Jacobian $J = \text{d} y/ \text{d} x$. From (\ref{trafo_general}) we get
\begin{equation}
p(x) = J\, \pi (y) \quad \text{and} \quad \frac{\partial}{\partial x} = J \, \frac{\partial} {\partial y}
\end{equation}
In order to obtain constant diffusion we have to require
\begin{equation}
J^2 D \equiv \tilde{D} = \text{const} \quad \to \quad \partial_y (J^2 D) = 0
\label{requirement}
\end{equation}
The transformed Fokker-Planck equation then reads
\begin{equation}
\partial_t \pi (y, t) = -2 \Omega^2 t \left[ \partial_y \tilde{F}(y) - \frac{\tilde{D}}{2}
\partial_y^2 \right] \pi (y,t )
\label{trafo_fokker_planck}
\end{equation}
with
\begin{equation}
\tilde{F} = J \left[ F - \frac{1}{2} \frac{\partial}{\partial y} \left( D J \right) \right]
\label{effective_force}
\end{equation}
Finally, if we set $\tilde{D} = 1$ it follows from (\ref{requirement}) that
\begin{equation}
J = \frac{\text{d} y}{\text{d} x} = \sqrt{\frac{1}{D(x)}}
\label{diff_eq}
\end{equation}
so that the transformation $x \to y(x)$ can in principle be explicitly calculated.

The solution of eq. (\ref{diff_eq}) can be expressed analytically in terms of various elliptic integrals over inverse trigonometric and hyperbolic functions. It is, however, not very illuminating. Luckily, in the range $x = 0 \dots 1/2$ the solution can excellently be fitted by a quadratic function and thus the inverse transformation also be determined
\begin{eqnarray}
y (x) &=& a_1 x + a_2 x^2 \nonumber \\
x (y) &=&  \frac{1}{2 a_2} \left( \sqrt{a_1^2 + 4 a_2 y} -a_1\right)
\label{trafo_explicit}
\end{eqnarray}
with $a_1 = 0.7074$ and $a_2 = 0.4169$. The effective force $\tilde{F} (y)$ can now explicitly be calculated.


\begin{thebibliography}{10}
\bibitem{footnote} We note that the Master Equation has a stationary solution despite the explicit appearance of the time variable on the right hand side.  This can be seen by transforming to a new time variable $\tau = \Omega t^2$ which removes the explicit time-dependence. The detailed balance condition for the transition coefficients ensures that a steady state is reached.
\end{thebibliography}
\end{document}